\documentclass[runningheads]{cl2emult}

\usepackage{makeidx}  
\usepackage{graphicx} 
\usepackage{subeqnar} 
\usepackage{multicol} 
\usepackage{cropmark} 
\usepackage{eso}      
\makeindex            

\begin{document}
\title*{Classification and Redshift Estimation\protect\newline
in Multi-Color Surveys}
\toctitle{Classification and Redshift Estimation\protect\newline
in Multi-Color Surveys}
%
%
\titlerunning{Classification and Redshift Estimation}
%
\author{Christian Wolf
\and Klaus Meisenheimer
\and Hermann-Josef R\"oser}
\authorrunning{Christian Wolf et al.}
%
%
\institute{Max-Planck-Institut f\"ur Astronomie,
K\"onigstuhl 17, 69117 Heidelberg, Germany}

\maketitle              

\begin{abstract}
We present a photometric method for identifying stars, galaxies and quasars in multi-color surveys and estimating multi-color redshifts for the extragalactic objects. We use a library of $> 65000$ color templates for comparison with observed objects. The method was originally developed for the Calar Alto Deep Imaging Survey (CADIS), but is now used in a variety of survey projects. We checked its performance by spectroscopy of CADIS objects, where it provides high reliability (6 mistakes among 151 objects with $R<24$), especially for the quasar selection, and redshifts accurate within $\sigma_z \approx 0.03$ for galaxies and $\sigma_z \approx 0.1$ for quasars. For an optimization of future surveys, a few model surveys are compared, which use the same amount of telescope time but different filter sets. In summary, medium-band surveys perform superior to broad-band surveys although they collect less photons. A full account of this work is already in print ([11,12]).
\end{abstract}

\section{Introduction}

The value of a multi-color survey is usually not simply limited by the depth of its imaging but by the depth of its classification used to select the objects of interest from the survey data. Here, we investigate what factors the depth of a successful classification depends on. Obviously, it depends in some way on the filter set used as well as on the depth of the imaging, but the optimum choices might vary according to the goal of the survey. 

If a survey aims at identifying only one type of object with characteristic colors, a tailored filter set can be designed. E. g., when looking for U-band dropouts [9], the {\it ugr} filter set is a very good choice. The performance of such a dropout survey depends mostly on the depth reached in the U-band, so the photon flux detection limit in U is the key figure. Also, number count studies are limited by the completeness limit in the filter of concern.

If all kinds of different objects are the subject of the survey, a filter set covering the entire spectral range accessible to the survey instrument is a natural choice. The SLOAN Digital Sky Survey [13] is presently the most ambitious project to provide a broad-band {\it ugriz} color database, on which the astronomical community might perform a large number of virtual surveys. 

Of course, it is possible to split the bands of a filter set even further into a larger number of medium-band filters. Surveys like CADIS with typically 10 to 20 filters are sampling the visual spectrum with a resolution almost comparable to that of low resolution imaging spectroscopy. Intuitively, one might expect that such surveys are very expensive in terms of telescope time while their particular value might not be obvious. This paper will show, that in fact medium-band surveys are a powerful and economically competitive alternative to broad-band surveys.

\section{Classification and Redshift Estimation in CADIS}

CADIS fostered the development of a scheme for spectral classification, that distinguishes {\it stars}, {\it galaxies}, {\it quasars} and {\it strange objects}. Simultaneously, it assigns multi-color redshifts to extragalactic objects, an idea dating back to 1962 [1]. Essentially, a likelihood-based algorithm compares the observed colors with a library of template colors. The probability of each single template to generate the colors of the observed object is calculated and a probability density function is obtained for each object class, dependent on the redshift and the spectral energy distributions (SED) contained in the library. From these functions, relative likelihoods for class memberships and estimates for redshift, SED types and their errors can all be derived (see [11] for all details).

The color template libraries are assembled from observed spectra by synthetic photometry performed on our filterset. As an input we used the stellar library by Pickles [8], the galaxy template spectra by Kinney et al. [5] and the QSO template by Francis et al. [3]. From this, we generated regular grids of quasar templates ranging in redshift within $0<z<6$ and having various continuum slopes and emission line equivalent widths. The quasar spectra are attenuated bluewards of the Lyman-$\alpha$ line by a redshift-dependent throughput function modelling the Lyman forest. Also, a grid of galaxy templates has been generated for $0<z<2$, and contains various spectral types from old populations to starbursts. 

Using this method, the object database of CADIS was classified and the result was checked by spectroscopy for 151 objects with $R<24$. An object with $R=24$ is typically detected at a 5-$\sigma$ level in most of the up to 17 filters. At $R<22$ the classification was almost free of mistakes with two mistakes among 103 objects (two Seyfert-1-QSOs were classified as normal galaxies). In the fainter regime of $R>22$, a morphological criterion was applied to all extended objects classifing them directly as galaxies and four classification mistakes were finally found among 48 objects. 

Multi-color redshifts were accurate to $\sigma_z \approx 0.03$ for 90\% of the galaxies at $R<24$ while the remaining 10\% are estimated significantly wrong. Among the quasars, 50\% were accurate to $\sigma_z \approx 0.1$ with another 50\% of complete mistakes. At the present time, the observed quasar color spectra are affected by intrinsic variability during the four years of CADIS observations needed to collect the data for all the bands. Once this effect is corrected for we expect improvements for the multi-color redshifts of the quasars (see [12] for details on the CADIS results).

\begin{table}
\centering
\setlength\tabcolsep{10pt}
\caption{Filters and 10-$\sigma$-magnitude limits for the three survey setups (A, B, C) compared with Monte-Carlo simulations. The I-band filter is a long wavelength passband filter with a cut-on wavelength roughly at 780\,nm. Its far-red sensitivity limit is given by the dropping quantum efficiency of the CCDs. All filters are installed in the Wildfield Imager at the 2.2m-MPG/ESO telescope at La Silla observatory. Setup A has actually been used for the deep WFI survey {\it COMBO-17}. \label{extime}}
\begin{tabular}{lcccc}
$\lambda_{cen}$/fwhm (nm) & name & $m_{lim,A}$ & $m_{lim,B}$ & $m_{lim,C}$ \\
\noalign{\smallskip}
\hline
\noalign{\smallskip}
364/38  	& U & 23.5 & 23.5  & 24.1  \\
456/99  	& B & 25.0 & 25.0  & 25.6  \\
540/89  	& V & 24.5 & 24.5  & 25.1  \\
652/162 	& R & 24.5 & 24.5  & 25.1  \\
850/150*	& I & 23.0 & 23.0  & 23.6  \\
\noalign{\smallskip}
420/30	& & 23.6 & 23.98 &   \\
462/14	& & 23.5 &       &   \\
485/31	& & 23.4 & 23.78 &   \\
518/16	& & 23.3 &       &   \\
571/25	& & 23.2 & 23.58 &   \\
604/21	& & 23.1 &       &   \\
646/27	& & 23.0 &       &   \\
696/20	& & 22.8 & 23.18 &   \\
753/18	& & 22.7 &       &   \\
815/20	& & 22.6 & 22.98 &   \\
856/14	& & 22.5 &       &   \\
914/27	& & 22.4 & 22.78 &   \\
\noalign{\smallskip}
\hline
\end{tabular}
\end{table}

\section{Model Surveys in Comparison}

Initially, it should be natural to assume that surveys with different filter sets show quite a different performance in terms of classification and redshift estimation. If a survey aims for  objects with very particular spectra, the filter set can certainly be tailored to this purpose. If the objects of interest span a whole range of spectral characteristics, it is not trivial to guess via analytic thinking which filter set performs best. Here, we present a comparison of three fundamentally different filter sets and show their performance for classification and redshift estimation (see [12] for all details).

The three model surveys spend the same total amount of exposure time on different filter sets, using the same instrument, telescope and observing site. We chose the Wild Field Imager (WFI) at the 2.2-m-MPG/ESO-telescope on La Silla as a testing ground, because it provides a unique, extensive set of filters ranging from several broad bands to a few dozen medium bands to choose from. Furthermore, the WFI is a designated survey instrument which is extensively used by the astronomical community.

The three modelled surveys, here called setup ``A'', ``B'' and ``C'', each spend 150\,ksec of exposure time distributed on the following filters (see also Tab.\,\ref{extime}): 

Setup A spends 50\,ksec on the five broad-band filters of the WFI (UBVRI) and 100\,ksec on twelve medium-band filters. Using ESO's exposure time calculator V2.3.1 for the WFI, we related exposure times to limiting magnitudes assuming a seeing of 1.4 arcsec, an airmass of 1.2, point source photometry and a night sky illuminated by a moon three days old. The exposure times are distributed such, that a quasar with a power-law continuum $f_\nu = \nu^\alpha$ and a spectral index of $\alpha = -0.6$ is observed with a uniform signal-to-noise ratio in all medium bands. As a result, the twelve medium bands each deliver a 10-$\sigma$ detection of an $R=23.0$-quasar. Setup B spends 50\,ksec on the same broad bands but concentrates the 100\,ksec for medium-band work on only six filters reaching a uniform 10-$\sigma$ detection of a $R=23.38$-quasar then. Setup C finally spends all 150\,ksec on the broad-band filters and omits the medium bands entirely. 

\begin{figure}
\centering
\includegraphics[width=1\textwidth]{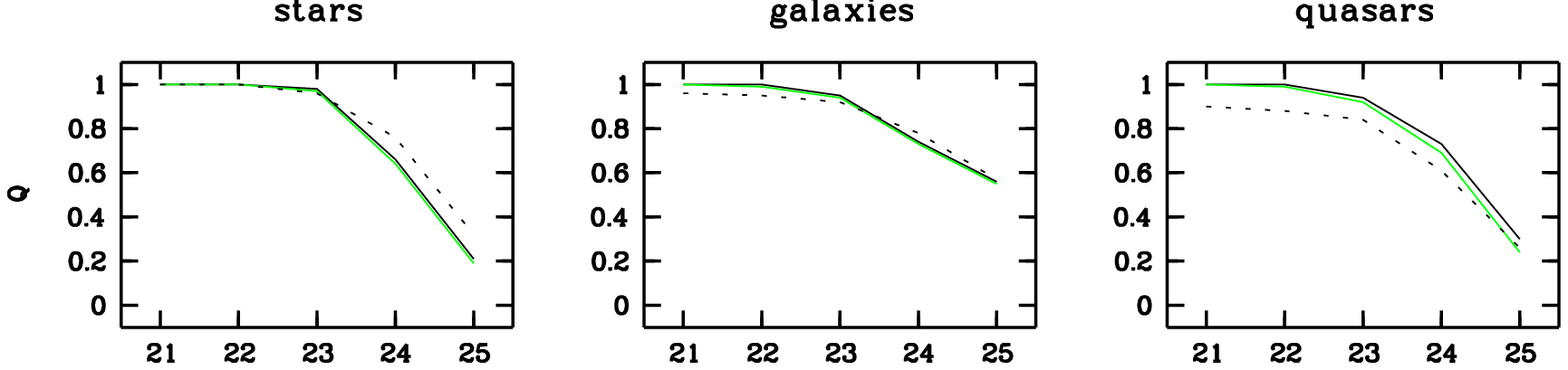}
\caption[ ]{Fraction Q of simulated objects which are correctly classified in the three different setups (solid line = setup A, grey line = setup B, dashed line = setup C) over the R-band magnitude in a 150\,ksec-survey at the WFI@2.2 (see Tab.\,1 for details). Basically, all setups are comparably successful (taken from [11]). \label{q3}}
\end{figure}

The resulting performance was evaluated in terms of the fraction of successfully classified objects as a function of class, magnitude and survey setup and also in terms of the achieved accuracy of the multi-color redshifts. As shown in Fig.\,\ref{q3}, all surveys perform rather equal in terms of class recovery. At $R=21\ldots 23$, the surveys work almost perfect and then degrade quickly with magnitude until becoming useless at $R>24$. Obviously, the setups with medium-band filters included (A, B) collect much less photons than the pure broad-band setup (C), but deliver a comparable classification performance. Therefore, the smaller number of detected photons is compensated by the increased information content per photon.

\begin{figure}
\centering
\includegraphics[width=1\textwidth]{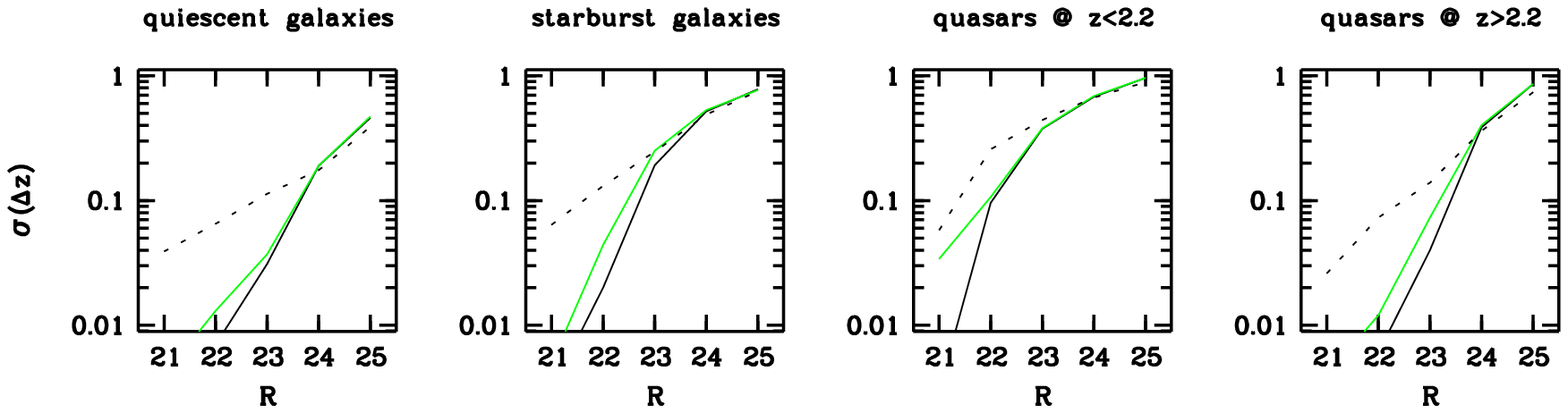}
\caption[ ]{Variance of redshift estimation error ($\Delta z = z_{mc}-z$) among simulated objects in the three different setups (lines as in Fig.\,1). Setups A and B provide the highest redshift resolution. Early type galaxies work better due to their higher continuum contrast at the 4000\,\AA-break. \label{sz22}}
\end{figure}

In contrast, the accuracy of the multi-color redshifts shows very significant differences between the setups (Fig.\,\ref{sz22}). In the bright regime at $R<22$, the redshift resolution is proportional to the wavelength resolution of the filter set, setting the medium-band surveys far ahead the broad-band survey. The information potential of medium-band filters, which do not average the object spectra as broadly as broad-band filters do, can fully be exploited and cause a clear advantage for the total performance of medium-band surveys. Only in the faint regime at $R>24$, where the redshift estimation is almost useless, the differences between the setups vanish.

These results on the performance are average figures for large samples. In this context it is important that pure broad-band surveys contain a few class ambiguities in multi-color space. After dropping the u-band from the SDSS filter set (which provides only a weak detection of faint objects anyway), an M2V star and a galaxy of type Sb at $z=0.5$ can not be distinguished by color, even if the relative calibration of the filters is accurate to 3\%. Similarly, a QSO at $z=3$ with $\alpha = -0.8$ displays the same colors as a G2V star. In contrast, the two medium-band filter sets can easily distinguish these pairs of objects to rather faint levels.

\section{Information Content of Multi-Color Surveys}

The rather equal performance of the different survey setups can be understood by analytically considering an infomation content for the surveys, given by the product of the following factors:
\begin{itemize} 
\item number of filters $n$
\item exposure per filter $t \propto n^{-1}$
\item filter width $\Delta\lambda \propto n^{-1}$
\item {\it information} per detected photon $\Phi \propto \Delta\lambda^{-1} \propto n$
\end{itemize}

The total information content yields $I_{tot} \propto n^0 = $const. This result is only valid under the ideal assumption that photon noise was the only source of error and the color libraries mimic true nature perfectly. But in a real survey there will be calibration errors limiting the achievable best accuracy to 3\% (equivalent to a 30-$\sigma$ detection) and also cosmic variance will cause observed spectra to deviate in a scattered manner from the library templates. In the bright regime, where performance is limited by these systematic effects and not by photon statistics, many noisier datapoints provide more information than fewer precise but inaccurate ones. So, under realistic circumstances the medium-band survey setups are likely to be more powerful than the pure broad-band setup even when being limited by equal telescope time.

\section{Ongoing Applications}

We currently work on a number of multi-color surveys, where filters and exposure times were chosen to match some primary survey strategy. Although, none of these might have been optimal choices in terms of a general classification, we used or intend to use our approach to extract class and redshift data on the objects contained. These surveys are:

\begin{enumerate}
\item The {\it Calar Alto Deep Imaging Survey} (CADIS): Four broad-band (BRJK$^\prime$) and twelve medium-band filters have mostly been chosen to match the needs of the emission line survey in CADIS, while some of them fill in gaps in the spectral coverage. The multi-color part of CADIS has been used to study the evolution of the galaxy luminosity function at $z=0.1\ldots 1.3$, to search for quasars at all visible redshifts and to use the observed faint stellar population to check models of the Galactic structure and the stellar luminosity function [4,6,7,10,12].

\item A lensing study of the galaxy cluster Abell 1689: Two broad-band and seven medium-band filters have been chosen to separate well between the cluster galaxies at $z \approx 0.19$ and the background population. The galaxy luminosity function in the background of the cluster is compared to a control field taken from CADIS, and the cluster mass is estimated from weak lensing effects on the apparent luminosities [2].

\item A WFI survey for {\it Classifying Objects by Medium-Band Observations with 17 filters} (COMBO-17): The filters (setup A from Tab.\,1) are chosen to provide a selection function and a redshift accuracy for quasars and galaxies, which is as independent of redshift as feasible. The data will be used to study the faint end of the quasar luminosity function at all accessible redshifts $z > 1$, galaxy-quasar correlation at $z<1.3$, as well as weak lensing effects in the cluster group Abell 901/2 and in the open galaxy field.

\item The {\it Sloan Digital Sky Survey} (SDSS): Five broad filters are used to span the entire range of available CCD sensitivity. We intend to apply our classification to get multi-color redshifts, to search for quasars and to separate stars from compact galaxies, where morphology data are not sufficient.
\end{enumerate}

\section{Summary}

We presented an innovative method that performs a multi-color classification and redshift estimation of astronomical objects in a unified approach. The method is essentially based on templates and evaluates the statistical consistency of a given measurement with a database of spectral knowledge, serving as a second, very crucial input to the algorithm. Spectroscopic confirmation within CADIS showed that the method works well and is of high practical relevance.

A simulation of a few model surveys with fundamentally different filter sets revealed that in practice surveys with many medium-band filters are superior to pure broad-band surveys for the purpose of classification and redshift estimation. Even when being limited by telescope time, the medium-band filter sets achieve at least the same working depth as the broad-band survey while providing much improved redshift resolution.

We conclude, that this method should be very suitable for many survey applications, which usually require only low spectral resolution and finite accuracy in the derivation of physical parameters, but aim for large samples to feed statistical studies and to search for rare and unusual objects. Of course, if a 100\% sure confirmation on the nature of an individual object is needed, or if you aim for high resolution studies, the method gives only a preselection of candidates.

\clearpage
\addcontentsline{toc}{section}{Index}
\flushbottom
\printindex

\end{document}